\documentclass[preprint,authoryear,12pt]{elsarticle}

\usepackage{graphicx}

\usepackage{rotating}

\usepackage{amssymb}
\usepackage{amsthm}

\journal{New Astronomy}

\begin{document}

\begin{frontmatter}



\title{Ion-cyclotron Resonance with Streaming Bi-Maxwellian Distribution}


\author[label1]{{S. Do\u{g}an}\corref{cor1}}
\ead{suzan.dogan@ege.edu.tr} \cortext[cor1]{Corresponding author}
 \address[label1]{University of Ege, Faculty of Science, Department of Astronomy and Space Sciences, Bornova, 35100, \.Izmir, Turkey}

\author[label1]{E. R. Pek\"{u}nl\"{u}}

\address{}

\begin{abstract}
We investigate the effect of bulk velocity of the solar wind on the
propagation characteristics of ion-cyclotron waves (ICWs). Our model
is based on the kinetic theory. We solve the Vlasov equation for O
VI ions and obtain the dispersion relation of ICWs. Refractive index
of the medium for a streaming bi-Maxwellian velocity distribution
proved to be higher than that of the bi-Maxwellian velocity
distribution. The bulk velocity of the solar polar coronal holes'
plasma increases the value of the refractive index by a factor of
1.5 (3) when the residual contribution is included (neglected). The
ratio of the refractive index of interplume lanes to the plume lanes
at the coronal base is also higher than we found for the
bi-Maxwellian velocity distribution, i.e. $k_{\rm interplume}/k_{\rm
plume}=2.5$.
\end{abstract}

\begin{keyword}
Sun: corona - solar wind - waves - acceleration of particles


\end{keyword}

\end{frontmatter}


\section{Introduction}
The UVCS/SOHO observations of plasma properties in the polar coronal
holes (PCH) revealed the existence of preferential ion heating with
$T_{i}\gg T_{p} > T_{e}$ (Kohl et al., 1997a; 1998). Ion to proton
kinetic temperature ratios are significantly higher than the ion to
proton mass ratios, i.e. $T_i/T_p>m_i/m_p$ (Kohl et al. 1999). The
UVCS measurements on PCH also provided clear signatures of
temperature anisotropies, with the temperature perpendicular to the
magnetic field being much larger than the temperature parallel to
the field. The temperature anisotropy, for example, is found to be
$T_{\perp}/T_{\parallel}\approx 10-100$ for O VI ions (Kohl et al.,
1997b; 1998; Cranmer et al. 1999). These observations led to
widespread interest in theoretical models of ion-cyclotron resonance
(ICR), because this mechanism naturally produces preferential
heating and temperature anisotropies (e.g. Dusenbery and Hollweg,
1981; Cranmer, 2001; Isenberg and Vasquez, 2009).

The ICR mechanism is a classical wave-particle interaction between
the left-hand polarized ion-cyclotron/Alfv$\acute{\rm e}$n waves and
the positive ions gyrating around the background magnetic field
lines. Resonance occurs when the Doppler-shifted wave frequency and
the ion cyclotron frequency are equal. When the resonance condition
is met, the electric field oscillations of the wave are no longer
\textquotedblleft felt" by the ion. Therefore, the ion sees a
constant electric field and can efficiently absorb energy from the
wave depending on the relative phase between its own velocity vector
and the electric field vector of the wave. The resulting effect of
this process is to increase the perpendicular velocity of the ion
(Isenberg and Vasquez, 2009; Cranmer 2009). Ions with higher
perpendicular energy diffuse into wider Larmor orbits. The
preferential perpendicular heating indicated by UVCS measurements
naturally leads to a preferential acceleration since the ions having
greater perpendicular velocities experience a higher mirror force,
$F_{\rm mf}(R)=-\mu\nabla B= -[(1/2) m_{\rm i}
v^{2}_{\bot}/B(R)]\nabla B $, in the radially decreasing magnetic
field of the coronal hole.

In a series of papers by Hollweg (1999a, b, c) the authors
investigated the resonant interactions with ion-cyclotron waves in
coronal holes in detail. Besides, Hollweg and Markovskii (2002)
offered a physical discussion of how the cyclotron resonances behave
when the waves propagate obliquely to the magnetic field. Vocks and
Marsch (2002) described a kinetic model which is based on
wave-particle interactions. Their model successfully explains the
preferential heating of heavy ions and the temperature anisotropies.
Cranmer (2000) investigated the dissipation of ion-cyclotron
resonant waves by taking more than 2000 ion species into account and
showed the effective damping ability of the minor ions.

There are also a number of theoretical models which are based on
fluid or magnetohydrodynamic approaches to the coronal heating and
the solar wind acceleration (e.g. Tu and Marsch, 1997; Suzuki and
Inutsuka, 2006; Matsumoto and Suzuki, 2012). The suggested fluid
models can successfully produce the required energy flux density to
heat the coronal holes without resonant interactions. However, the
kinetic approach has an advantage of distinguishing the plasma
particles in a more definitive way in accordance with the
collisionless nature of PCH (Cranmer, 2009). A kinetic description
is required in order to account fully for the observed microscopic
details of the solar wind plasma (Marsch, 1991). Since the ICR
process successfully explains the details revealed by the
observations, i.e. the presence of temperature anisotropy of minor
ion species and preferential heating, it has been proposed as the
most favourable heating mechanism of the solar corona.

Over the last few years, there has been an enormous increase in the
number of observations on ubiquitous waves in the solar corona.
Tomczyk et al. (2007) detected MHD waves by using Coronal
Multi-Channel Polarimeter (CoMP).  Their estimate of the energy
carried by the spatially resolved waves indicates that the waves are
too weak to heat the solar corona; however they also proposed that
the unresolved MHD waves can carry enough energy to heat the corona.
Recently, McIntosh et al. (2011) measured the amplitudes, periods
and phase speeds of outward-propagating Alfv\'{e}nic motions in the
transition region and the corona. They reported that the observed
waves are energetic enough to accelerate the fast solar wind and
heat the quiet corona. Besides, the observations of Hahn et al.
(2012) provide a new evidence of wave damping at unexpectedly low
heights in a polar coronal hole (PCH). They estimate that the
dissipated energy may account for a large fraction (up to about
70\%) of that required to heat the coronal hole and accelerate the
fast solar wind.

In Do\u{g}an \& Pek\"{u}nl\"{u} (2012, hereafter Paper I) we studied
the effect of the plume and interplume lanes (PIPL) of PCHs on the
interaction between the O VI ions and the ion-cyclotron waves
(ICWs). In Paper I as well as in the present one, the gradients of
number density and temperature along and perpendicular to the
magnetic field lines were considered. It was shown in Paper I that
the resonance process in the interplume lanes is much more effective
than in the plumes. Nevertheless, we should mention that more
efficient dissipation may not necessarily lead to a higher wind
flow. It is discussed by some authors that a stronger heating rate
is required to produce the denser wind flows in the plume regions
where the bulk speed is observed to be slower (see e.g. Wang, 1994;
Pinto et al. 2009).

The present investigation is the extension of Paper I wherein we
assumed that O VI ions have a bi-Maxwellian velocity distribution
function in PCHs. In this study, we assume that the streaming
bi-Maxwellian velocity distribution function for O VI ions would be
physically more realistic than that of bi-Maxwellian, since the
former will take into consideration the bulk velocity of the solar
wind. The paper is structured as follows: In section 2, we briefly
summarize the plasma properties of the PCHs and obtain the
dispersion relation of the ICWs solving the Vlasov equation for O VI
ions. We also compare the new results with those found in Paper I.
We present our conclusions in Section 3.

\section{Model}
\subsection{Plasma properties of PCH}
Properties of the PCH plasma were explained in Paper I in greater
detail (see section 2 of Paper I and references therein). Here, we
only summarize the radial dependences of effective temperature,
number density, the PCH magnetic field and the bulk velocity of O VI
ions, respectively. By using the empirical model of Cranmer et al.
(1999), Banerjee et al. (2000) derived a best-fit function for the O
VI line width which is valid in the range 1.5-3.5 $\rm R_{\odot}$.
When we convert their Eq. (3) into the effective temperature, we get
the radial dependence of $T_{\rm eff}$ as below:

\begin{equation}
T_{\rm{eff}}(R)=4.02 \times 10^{7}R^{2}+1.25 \times 10^{7}R-9.76
\times 10^{7} \,\,\, \rm K.
\end{equation}
where R is the dimensionless radial distance, i.e., R=r/$\rm
R_{\odot}$. By using the polarized white light and line ratio
measurements in northern PCH, Esser et al. (1999) derived an
analytical expression for the electron number density:
\begin{equation}
N_{\rm{e}}(R) =2.494\times 10^{6}R^{-3.76}+1.034\times
10^{7}R^{-9.64}+3.711\times 10^{8}R^{-16.86}\,\,\, \rm cm^{-3}.
\end{equation}

Radial dependence of PCH magnetic field is given by Hollweg (1999a)
as below:
\begin{equation}
B=1.5(f _{\rm max}-1)R^{-3.5}+1.5 R^{-2} \,\,\rm Gauss
\end{equation}
where $f_{\rm max}=9$. The radial dependence of O VI cyclotron
frequency ($\omega_{\rm c}=q_{\rm i}B/m_{\rm i}c$) will be
calculated by using Eq. (3).

Verdini et al. (2012) present a solar wind speed profile in the
radial direction. By using their Fig. 1, we express the radial
dependence of the wind speed with the equation given below,
\begin{equation}
\emph{u}=5R^3-45R^2+140R+150 \,\,\rm kms^{-1}.
\end{equation}
This relation is valid in the distance range 1.5-3.5 R. We should
mention that the solar wind speed profile strongly depends on the
magnetic field profile assumed in the wind model of Verdini et al.
(2012). The magnetic field profiles of Verdini et al. (2012) and
Hollweg (1999a) overlap quite a lot with appropriate values of free
parameters. Nevertheless, our calculation may include an imprecision
resulting from the possible inconsistency of two profiles. An
interested reader may refer to Pinto et al. (2009) and Grappin et
al. (2010) for alternative profiles.

We also consider the PIPL structure of PCH, the physical parameters
of which display gradients along and perpendicular direction to the
external magnetic field. Wilhelm et al. (1998) report that the
effective temperature of O VI ions in the interplume lanes are about
30\% higher than that of plumes. By using this observational result
Devlen $\&$ Pek{\"u}nl{\"u} (2010) formulate the temperature
structure of PCH in two dimensions as below:

\begin{equation}
T_{\rm{eff}}(R,x)= T_{\rm{eff}}(R)+ 0.3T_{\rm{eff}}(R)\sin^{2}(\frac {2\pi}{\lambda} x)
\end{equation}
where \emph{x} is the direction perpendicular to \emph{R} and
$\lambda$ is the expansion rate of the widths of PIPL in PCH in
arcseconds, i.e., $\lambda=92''.16R$. The factor 0.3 comes from the
temperature difference between the plume and interplume lane. The
number densities of electrons in plumes appear to be 10\% higher
than that of interplume lanes (Kohl et al., 1997a). When we consider
this difference, we can express the number density of O VI ions in
(R,x) plane as below:
 \begin{equation}
N_{\rm{i}}(R,x)=fN_{\rm{p}}^{\rm{PL}}(R)[1-0.1\sin^{2}(2\pi x/\lambda)] \,\,\, \rm cm^{-3}
\end{equation}
where \emph{f} is the mean value of O VI number density
($1.52\times10^{-6}N_{\rm p}$) as given by Cranmer et al. (2008).
The factor 0.1 comes from the difference of the number densities
between plumes and interplumes. We will use these observational
results to solve the Vlasov equation in the next subsection.

\subsection{Solution of the Vlasov Equation}
It is appropriate to use Vlasov equation to describe the
collisionless plasma like coronal holes in time intervals shorter
than the interspecies collision time. In this method, the
first-order perturbations to the velocity distribution function is
calculated in coordinates that follow the unperturbed trajectory of
the particles. Knowledge of the perturbed velocity in terms of the
first-order electric field allows us to calculate the current
density and the dielectric tensor. Substitution of dielectric tensor
into wave equation gives the dispersion relation for the ICWs (Stix,
1992). We follow the same procedure as we did in Paper I and adopt
the space time variation of the perturbed quantities as
$\exp[i(\emph{\textbf{k}}\cdot\emph{\textbf{r}}-\omega t)]$. Wave
equation is given (e.g. Stix, 1962, 1992) as below,

\begin{equation}
\textbf{k}\times \left(\textbf{k} \times
\textbf{E}\right)+\frac{\omega^{2}}{c^{2}} \kappa \cdot \textbf{E}=0
\end{equation}
\\
where $\kappa$ is the dielectric tensor and conventionally is
derived from the Vlasov equation. We again assume that both the
Lorentz force,
$q[\textbf{E}+\frac{1}{c}(\textbf{v}\times\textbf{B}_{0})]$ and the
pressure gradient force, $-\nabla p$ act upon O VI ions, where
symbols throughout this paper are the same as in Paper I and have
their usual meanings. With these assumptions, the quasi-linearized
Vlasov equation is written as,

\begin{equation}
\begin{array}{l}\frac{{\it d}\, {\it f}_{1}}{{\it d}\, {\it t}} =\frac{\partial \, {\it f}_{1}}{\partial \, {\it t}} +\textbf{v}\cdot \frac{\partial \, {\it f}_{1} }{\partial \, \textbf{R}} +\frac{{\it q}_{\rm i}}{{\it m}_{\rm i}} \left(\textbf{E}_{1} +\frac {1}{c}(\textbf{v}\times \textbf{B}_{0}) \right)\cdot \frac{\partial \, {\it f}_{1}}{\partial \, \textbf{v}} = \bigg[ -\frac{{\it q}_{\rm i}}{{\it m}_{\rm i} } \left(\textbf{E}_{1} +\frac {1}{c}(\textbf{v}\times \textbf{B}_{1}) \right) \\
\\
+\frac{{\it k}_{ \rm B}}{n_{0}m_{\rm i}} \left(T_{\rm eff}^{\rm \xi}\frac{\partial n_{\rm 0}}{\partial \textbf{R}}+T_{\rm eff}^{\rm \xi}\frac{\partial n_{0}}{\partial \textbf{x}}+
n_{0}\frac{\partial T_{\rm eff}^{\rm \xi}}{\partial \textbf{R}}+n_{0}\frac{\partial T_{\rm eff}^{\rm \xi}}{\partial \textbf{x}}\right) \bigg]\cdot \frac{\partial \, {\it f}_{0} }{\partial \textbf{v}}. \end{array}
\end{equation}
where $f_{0}$ and $f_{1}$ are the unperturbed and perturbed parts of
the velocity distribution function and $\textbf{E}_{1}$ and
$\textbf{B}_{1}$ are the wave electric and magnetic fields,
respectively. The time derivative is taken along the unperturbed
trajectories in phase - space of O VI ions. Schmidt (1979) gives
Left Circularly Polarized (LCP) wave interacting with the perturbed
velocity function of ions as,
\begin{equation}
\begin{array}{l}{f_{\rm L}=}
 {-\frac{q_{\rm i} }{m_{\rm i} } \left[\left(1-\frac{v_{\parallel} k}{\omega } \right)\frac{\partial \, f_{0} }{\partial \, v_{\bot } } +\frac{k\, v_{\bot } }{\omega } \frac{\partial \, f_{0} }{\partial \, v_{\parallel} } \right]E_{\rm x} \exp \, \left(i\theta _{0} \right)\frac{1-\exp \phi}{i(k\, v_{\parallel} -\omega + \omega _{c} )} +\nabla p_{1}}
 \end{array}
\end{equation}
\\
where $\nabla p_{1}$ is
\begin{equation}
 \nabla p_{1}=\frac{{\it k}_{ \rm B}}{n_{0}m_{\rm i}} \left(T_{\rm eff}^{\xi}\frac{\partial n_{0}}{\partial \textbf{R}}+T_{\rm eff}^{\xi}\frac{\partial n_{0}}{\partial \textbf{x}}+
n_{0}\frac{\partial T_{\rm eff}^{\xi}}{\partial \textbf{R}}+n_{0}\frac{\partial T_{\rm eff}^{\xi}}{\partial \textbf{x}}\right)
\end{equation}
\\
$f_{1}$ produces a current, x and y components of which are given in
Paper I. Using these currents one can derive the plasma dielectric
tensor for LCP wave as:

\begin{equation}
\begin{array}{l} { \kappa_{L} =1+4\pi \frac{J_{\rm x} /E_{\rm x} }{i\, \omega \ } =} {1-
 \frac{4\pi^{2} \,q_{\rm i}^{2}  }{im_{\rm i}\ \omega ^{2} } \int _{-\infty }^{+\infty }d\, v_{\parallel} \int _{0}^{\infty }\frac{\left(\omega -k\, v_{\parallel} \right)\, \left(\partial \, f_{0} /\partial \, v_{\bot } \right)+k\, v_{\bot } \left(\partial \, f_{0} /\partial \, v_{\parallel} \right)}{k\, v_{\parallel} -\omega + \omega _{c} } (1-\exp \phi ) v_{\bot }^{2} \, d\, v_{\bot }   } \\
\\
+ {\frac{4\pi^{2} q_{i} }{ \omega E_{\rm x} } (t-t_{0} )\int _{0}^{\infty } \nabla p_{1} v_{\bot }^{2} \, dv_{\bot } \, dv_{\parallel} }. \end{array}
\end{equation}
\\
We will obtain the dispersion relation for LCP wave by substituting
Eq. (11) into $\kappa_{L}=n^{2}=c^{2}k^{2}/\omega^{2}$ where n is
the refractive index of the medium for ICWs.

Observations revealed that PCHs' collisionless and streaming
plasma displays temperature anisotropies. Bearing these
observational facts in mind, we proceed our investigation with a
\emph{streaming bi-Maxwellian} velocity distribution function for O
VI ions,
\begin{equation}
f_{0} =n_{\rm i} \, \alpha _{\bot }^{2} \alpha _{\parallel} \, \pi
^{-3/2} \exp \left[-\left(\alpha _{\bot }^{2} v_{\bot }^{2} +\alpha
_{\parallel}^{2} (v_{\parallel}-u)^{2} \right)\right]
\end{equation}
\\
where \textbf{\emph{u}} represents the overall bulk velocity of the
PCH plasma and $\alpha_{\bot}=(2k_{\rm B}T_{\bot}/m_{i})^{-1/2}$ and
$\alpha_{\parallel}=(2k_{\rm B}T_{\parallel}/m_{i})^{-1/2}$ are the
inverse of the most probable speeds in the perpendicular and
parallel direction to the external magnetic field, respectively. The
velocity derivatives of $f_0$ which will be substituted in Eq. (11)
are obtained from Eq. (12) and finally the contribution of the
principle integral to the dispersion relation of LCP wave is found
as below,
\begin{equation}
\begin{array}{l} {k^{2} \, \left(c^{2} -\frac{\omega ^{2} }{k^{2} }
\right)-\frac{i\omega _{p}^{2} \, \omega }{\left(\omega -\omega _{c}
\right)} -\frac{i\omega _{p}^{2} \, ku}{\, \left(\omega - \omega
_{c} \right)} \left[\frac{T_{\bot } }{T_{//} } \left(\frac{ku}
{(\omega -\omega _{c} )}
+2\right)-\frac{\omega }{(\omega -\omega _{c} )} -1\right]} \\ {} \\
{-\frac{i\omega _{p}^{2} k^{2} \left(1+2\alpha _{//}^{2} u^{2}
\right)}{2\alpha _{//}^{2} \left(\omega -\omega _{c} \right)^{2} }
\left(\frac{\omega \, } {\left(\omega -\omega _{c} \right)^{} }
-\frac{T_{\bot } }{T_{//} } +1\right)+\frac{\pi^{2}\omega {\it
q}_{i}\, L }{{\it E}_{\rm x} v_{A}}\nabla p_{1}^{\rm r}=0}
\end{array}
\end{equation}
\\
where $\omega_{p}=(4\pi n_{OVI}q_{i}^{2}/m_{i})^{1/2}$ is the plasma
frequency for the O VI ions and $\nabla p_{1}^{r}$ is the term
designating rather lengthy pressure gradient in its reduced form:
\begin{equation}
\nabla p_{1}^{\rm r}=\frac{{\it k}_{ \rm B}}{m_{\rm i}}\left[{
\left(T_{\rm eff}^{\xi}\frac{\partial n_{0}}{\partial
R}+n_{0}\frac{\partial T_{\rm eff}^{\xi}}{\partial R}\right)
\left(\frac{T_{\bot}}{T_{\parallel}}\right)^{1/2}}+2\left(T_{\rm
eff}^{\xi}\frac{\partial n_{0}}{\partial x}+n_{0}\frac{\partial
T_{\rm eff}^{\xi}}{\partial x}\right)\right].
\end{equation}

The residual contribution is found as,

\begin{equation}
\Re=\frac{2\sqrt{\pi } \omega _{p}^{2} }{k} \, \alpha _{//}
\left[\omega _{c} +(\omega -\omega _{c} -uk)\frac{T_{\bot } }{T_{//}
} \right]\exp \left[-\alpha _{//} ^{2} \left(\frac{\omega -\omega
_{c} }{k} -u\right)^{2} \right].
\end{equation}
\\
The dispersion relation is the combination of the principle and the
residual contribution given in Eqs (13) and (15) respectively:
\begin{equation}
\begin{array}{l} {k^{2} \, \left(c^{2} -\frac{\omega ^{2} }{k^{2} } \right)-\frac{i\omega _{p}^{2} \, }{\left(\omega -\omega _{c} \right)} \left[\omega +ku\frac{T_{\bot } }{T_{//} } +ku\Psi +\frac{k^{2} \left(1+2\alpha _{//}^{2} u^{2} \right)}{2\alpha _{//}^{2} \left(\omega -\omega _{c} \right)^{} } \Pi \right]} \\ {} \\ +\frac{\pi^{2}\omega {\it
q}_{i}\, L }{{\it E}_{\rm x} v_{A}}\nabla p_{1}^{\rm
r}{-\frac{2\sqrt{\pi } \omega _{p}^{2} }{k} \, \alpha _{//} \Phi\exp
\left[-\alpha _{//} ^{2} \left(\frac{\omega -\omega _{c} }{k}
-u\right)^{2} \right]=0} \end{array}
\end{equation}\\
where $\Psi$, $\Pi$ and $\Phi$ are
\begin{equation}
\begin{array}{l} {\Psi=\left[\frac{2T_{\bot } }{T_{//} } -\frac{\omega }{(\omega -\omega _{c} )} -1\right]}\\
\\
{\Pi=\left[\frac{\omega \, }{\left(\omega -\omega _{c} \right)^{} }
-\frac{T_{\bot } }{T_{//} } +1\right]} \\
\\
{\Phi=\left[\omega _{c} (\omega -\omega _{c})^2 +\frac{T_{\bot }
}{T_{//} } \left( \omega^3 - 3 \omega^2 \omega_c +3 \omega
\omega_c^2- \omega_c^3\right) \right]}.
\end{array}
\end{equation}\\

If we expand the exponential factor appearing in the last term of
the Eq. (16) into Taylor series, we get the \emph{Dispersion
Relation I} (DR I) as below,

\begin{equation}
\begin{array}{l} {k^{5} \left[c^{2} -\frac{i\omega_p^2 u^2}{(\omega-\omega_c)^2}
\frac{T_{\bot} }{T_{\parallel }}-\frac{i\omega _{p}^{2}
\left(1+2\alpha _{//}^{2} u^{2} \right)}{2\alpha _{//}^{2}
\left(\omega -\omega _{c} \right)^{2} } \Pi \right]-k^{4} \frac{{\it
i}\omega _{p}^{2} \, u}{\, \left(\omega -\omega _{c} \right)} \Psi}
\\ {}
\\
{-k^{3} \left[\omega ^{2} +\frac{{\it i}\omega _{p}^{2} \, \omega
}{\left(\omega -\omega _{c} \right)} -\frac{\pi^{2}\omega {\it
q}_{{\it i}} \, L } {{\it E}_{{\rm x}} v_{A} }\nabla p_{1}^{\rm
r}+2\sqrt{\pi}\omega _{p}^{2} \, \alpha _{//}^3 u \frac{T_{\bot}
}{T_{\parallel }}\right]} \\ {}
\\ {-2\sqrt{\pi } \omega _{p}^{2} \, \alpha _{//} k^{2}
\left[\omega_c + \omega_c \alpha _{//}^2 u^2+\left(\omega -\omega
_{c} \right)\frac{T_{\bot} }{T_{\parallel }}\left(1-3\alpha
_{//}^{2} u^{2} \right)\right]}\\
{}\\ {+{\it 2}\sqrt{\pi } \omega _{p}^{2} \, \alpha _{_{//} }^{}
{\it k}\left[3u\alpha _{//}^2 \frac{T_{\bot} }{T_{\parallel
}}\left(\omega -\omega _{c} \right)-2u\omega_c \alpha
_{//}^2\left(\omega +\omega _{c} \right) \right]}  {{\it
+2}\sqrt{\pi } \omega _{p}^{2} \, \alpha _{_{//} }^{3}
 { \Phi}=0}
\end{array}
\end{equation}

\begin{figure}
\centering
\includegraphics[angle=270,scale=0.6]{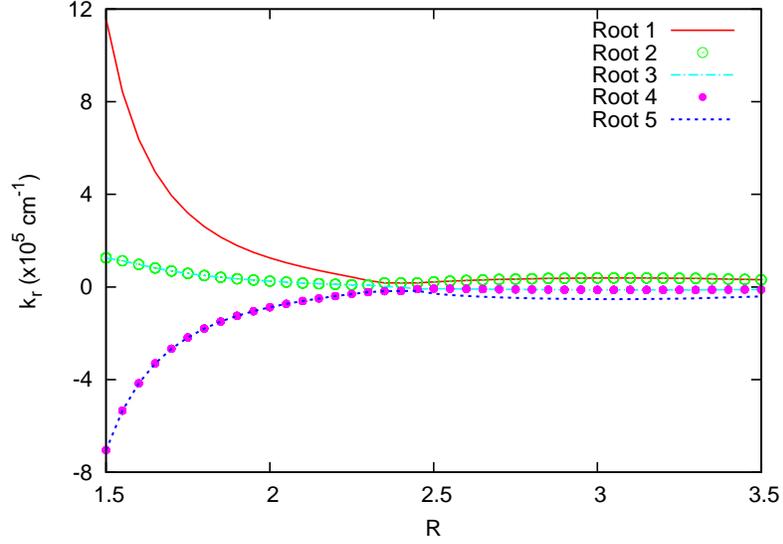}
\caption{All five roots of the fifth order \emph{DRI} given by Eq.
(18). The graph is for ICWs with a frequency 2500 $\rm rads^{-1}$.}
\end{figure}

\begin{figure}
   \centering
   \includegraphics[angle=270,scale=0.6]{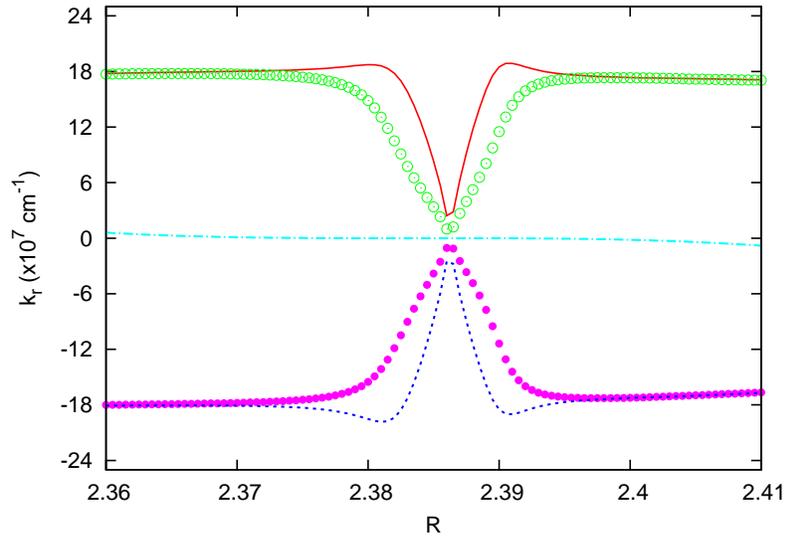}
 \caption{Fig. 1 is zoomed in the reflection region. Root 3 shows reflection at R=2.385. The same symbols are used with Fig. 1. }
    \end{figure}

\begin{figure}
   \centering
   \includegraphics[angle=270,scale=0.7]{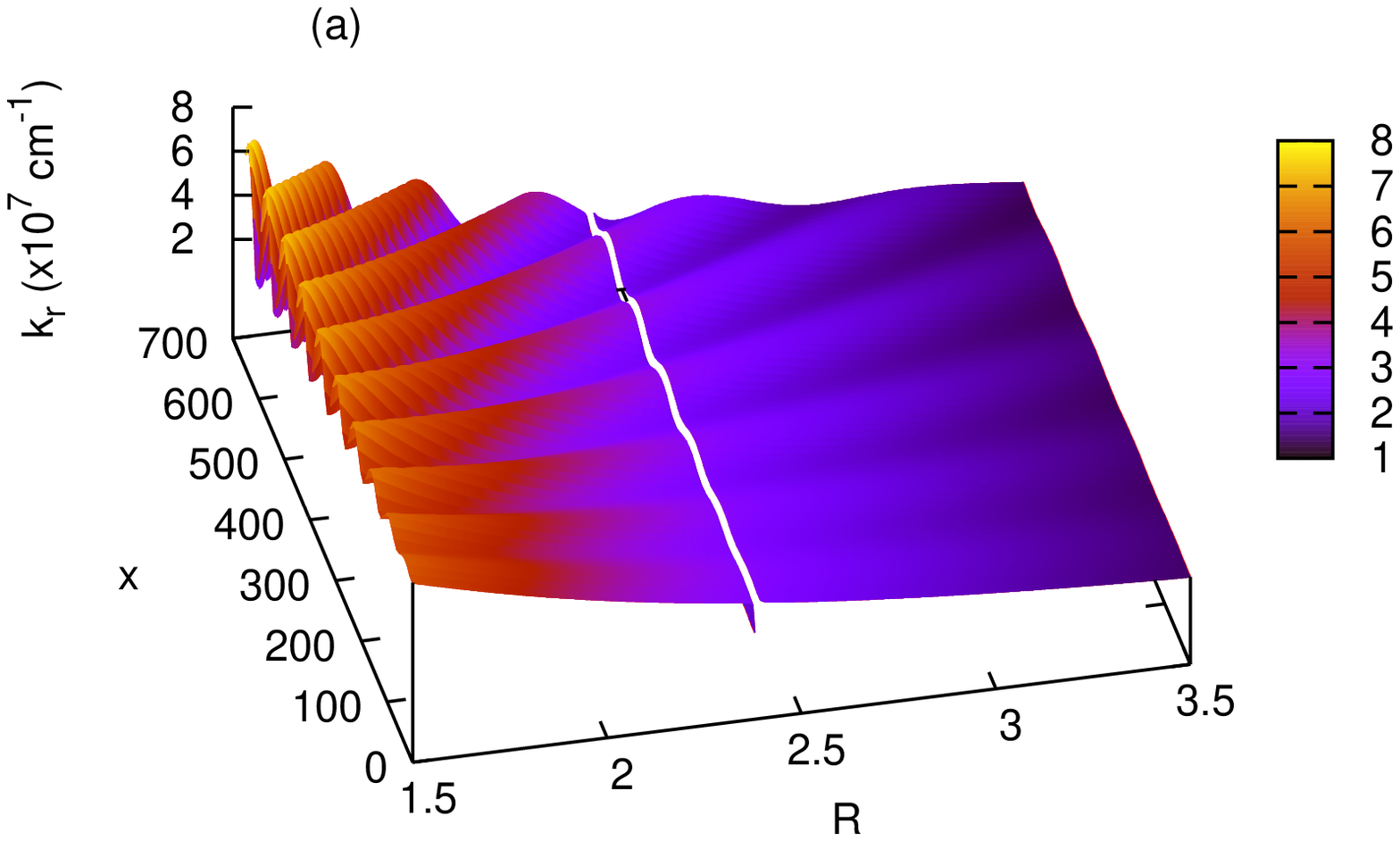}
   \includegraphics[angle=270,scale=0.7]{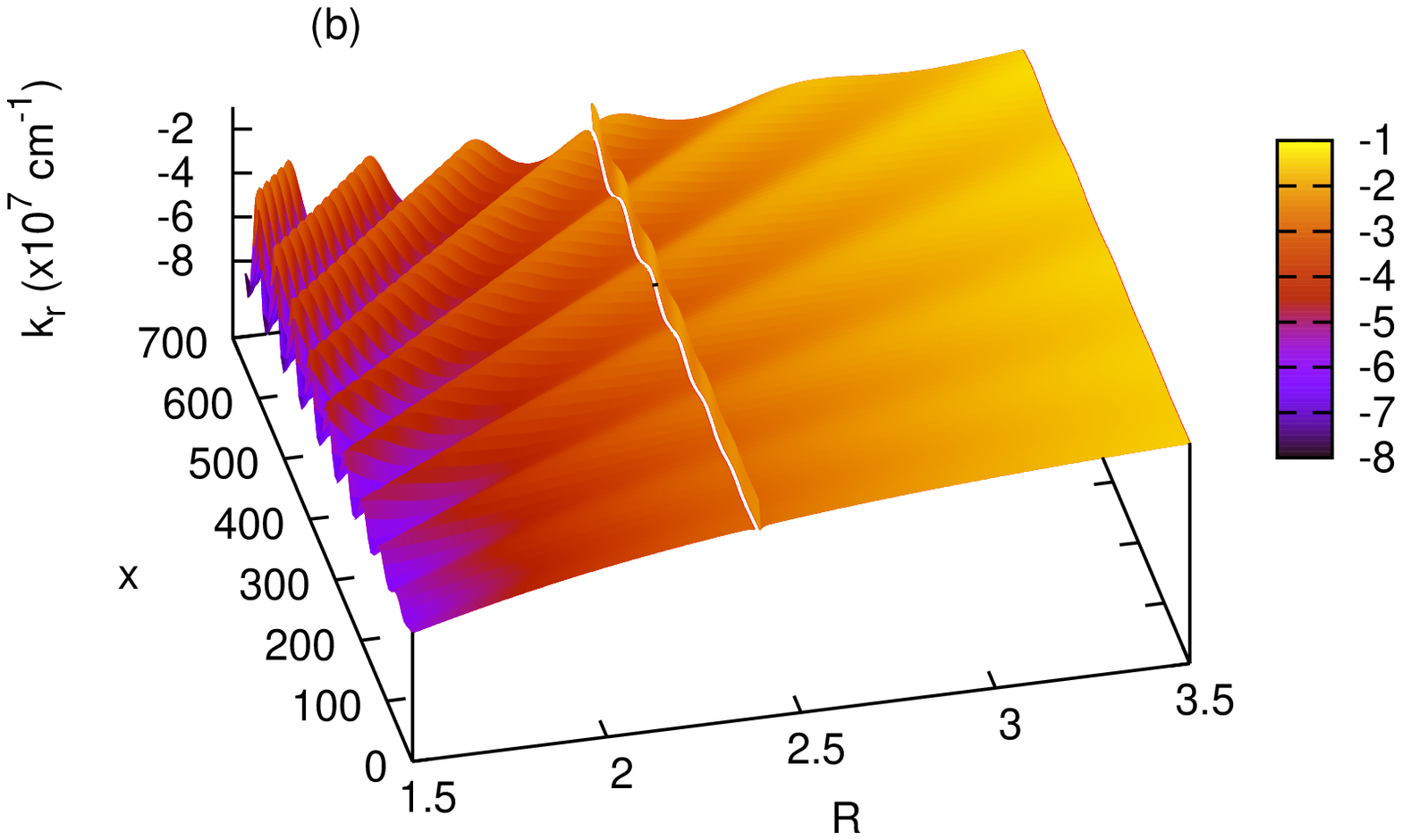}
\caption{Two roots of DR II. (a) First root of \emph{DRII}. This
root represents the backward propagating mode, (b) Second root of DR
II. This root represents the forward propagating mode. Both roots
reveal resonance at R=2.38. The higher values of $k_r$ in interplume
lanes point to more effective resonance process. The graphs are for
ICWs with a frequency 2500 $\rm rads^{-1}$.}
    \end{figure}

The real parts of all five roots of DR I (Eq. 18) are presented in
Fig. 1 in the radial distance range 1.5 R-3.5 R. According to
Shevchenko's criterion (2007), that is, $Im k^{2}=2k_{r} k_{i}
\gtrless 0$ where the upper (lower) sign is for the backward
(forward) wave, roots 1, 2, 3 and 5 represent the backward
propagating modes and root 4 is the forward propagating mode. It is
shown in Fig. 1 that all roots with a frequency of 2500 rads$^{-1}$
merge at about 2.38 R similar to the Fig. 2 of Paper I. This
location is the site where ion cyclotron resonance and cut-offs take
place in a small range of distance. Here we should remind the reader
that the field function of the wave is assumed of the form,
 $exp[i(\textbf{k}\cdot \textbf{r}-\omega t)]=exp[i(\textbf{k}_{r} \cdot \textbf{r}-\omega t)]exp(-\textbf{k}_{i}\cdot \textbf{r})$.
 Therefore, the situation where $k_r \rightarrow \infty $ corresponds to the refractive index going to infinity and implies a resonance.
  The situation where $k_r = 0 $ implies a cutoff and corresponds to a wave reflection. If the solution of the dispersion relation yields
  $ k_{i} > 0 $ then the wave damps in the $\textbf{r}$-direction. Differing from our previous results, the real part of the
wavenumber of the third mode becomes zero, corresponding to a
reflection at R=2.385. In going through cutoff, a transition is made
from a region of propagation to a region of evanescence. This result
is consistent with the positive values of the imaginary part of this
root after reflection. The amplitude of the mode shows a decay
through the radial distance indicating that the medium is not an
idealized reflector. No matter how small the decrease in wave
amplitude is, the dissipated wave energy is transferred to the O VI
ions (Melrose \& McPhedron, 1991).

The other difference from previous results is the value of the
wavenumbers. The wavenumbers are found to be, on average, 1.5 times
higher than those found in Paper I. The inclusion of the bulk
velocity of the PCHs' plasma increases the value of the refractive
index of the medium. In other words, the phase velocity of the ICWs
are found to be slower when the bulk velocity is taken into
consideration.

The reflection region where the refractive index is equal to zero
for root 3 is zoomed in Fig. 2. This wave attenuates spatially after
reflection. Fig. 2 also shows the modulational behaviour of the root
1, 2, 4 and 5 in the same reflection region. These modes approach
zero at R=2.385, however they do not show reflection. We should also
mention that the inclusion of the gradients in the x-direction
through the PIPL structure does not alter the solution of DR I. The
refractive index is found to be a function of radial distance (R)
only. We will show the effect of PIPL structure on the refractive
index in the solution of DR II.

The highest value of the residual contribution is found to be
$10^{-12}$ and this value is negligibly small compared to the values
of the rest of the terms which range between $10^6$ - $10^{13}$ in
the Eq. (16). If we neglect the residual contribution in the
dispersion relation given by Eq. (16) then Eq. (18) is reduced to DR
II given below:
\begin{equation}
{\it k}^{2} \left[{\it c}^{2} -\frac{\it i \omega_{p}^{2}u^2}
{(\omega-\omega_c)^2}\frac{{\it T}_{\bot } }{{\it T}_{{\it \parallel
}} }-\frac{i\omega _{p}^{2} \left(1+2\alpha _{//}^{2} u^{2}
\right)}{2\alpha _{//}^{2} \left(\omega -\omega _{c} \right)^{2} }
\Pi \right]-\frac{{\it i}\omega _{p}^{2} \, {\it ku}}{\,
\left(\omega -\omega _{c} \right)} \Psi-\frac{{\it i}\omega _{p}^{2}
\, \omega }{\left(\omega -\omega _{c} \right)} -\omega ^{2}+
\frac{\pi^{2}\omega {\it q}_{{\rm i}} \, L } {{\it E}_{{\rm x}}
v_{A} }\nabla p_{1}^{\rm r} =0
\end{equation}

The graphical solution of DR II is given in Fig. 3. As we mentioned
above, this solution reveals an infinity in the refractive index
indicating that a resonance occurs at R =2.38. The refractive index
is, on average, \textbf{3} times higher than those found in Paper I.
Including the bulk velocity of the solar wind clearly increases the
value of the refractive index. This result again implies that the
phase velocity of the ICWs becomes slower when the bulk velocity is
taken into consideration. Besides, the effect of the PIPL structure
is clearly seen in Figs. 3a-b. The real part of the wavenumbers
reveal a change both in R and x-direction. The crests correspond to
the interplume lanes and the troughs to the plume lanes. The
refractive index in the interplume lanes are found to be about 2.5
times higher than the ones in the plumes at the coronal base. We may
anticipate that the refractive index of the interplume lanes is
readily going to infinity indicating that the resonance process in
the interplume lanes is more effective than in the plumes.

\section{Conclusions}

In this paper, the effect of the PIPL structure of PCH is
re-discussed with the assumption of streaming bi-Maxwellian velocity
distribution which is expected to be more realistic for the PCH
plasma. The wavenumbers of ICWs for a streaming bi-Maxwellian
velocity distribution is found to be 1.5-3 times higher than that of
the bi-Maxwellian velocity distribution. Since the wave frequency we
assumed, i.e. 2500 $\rm rads^{-1}$, is the same in both
investigation, the higher wavenumber implies that the phase velocity
of the ICWs becomes slower when the bulk velocity is taken into
consideration. Wave spectrum in PCHs determines the net acceleration
to be acquired by the O VI ions having a streaming bi-Maxwellian
velocity distribution.

The refractive index is 2.5 times higher in interplume lanes than
the one in plume lanes. This ratio is slightly higher than we found
for the bi-Maxwellian distribution assumption. Higher refractive
index may imply that the resonance process is more effective in
interplume regions. However, it is yet to be discussed whether a
stronger heating is needed in order to produce faster flows of the
interplume lanes (see Pinto et al. 2009). If more efficient
resonance process means a higher wind speed, than our result is
consistent with the observations which reveal that the source of the
fast solar wind is interplume lanes.

Let us suppose that an O VI ion resonates with a wave of a certain
frequency, gains energy and escapes from the potential well of the
wave. If there is a wave with a higher frequency in the surrounding,
O VI ions come into resonance with it too and acquire more energy.
In order stochastic acceleration to come about there should be a
wide band of waves in PCHs. McIntosh et al. (2011) measured the
phase speeds, amplitudes and periods of the transverse waves both in
the quiet corona and PCHs. As to whether they also determined the
frequency band of the waves is unknown to the authors of the present
investigation.

In ICR process energy of the waves are transferred to the
perpendicular degree of freedom of the O VI ions in their helical
trajectories around magnetic field lines. Those particles the
velocity of which are equal to the phase velocity ($v = \omega /k$)
are readily trapped in the potential wells of the waves. Particles
with velocities slightly higher than the phase velocity of the wave
attempt to escape from the potential well of the wave, fail in this
attempt fall back and set in a oscillatory motion in the potential
well. These kind of particles, because of their initial velocity
were higher than the phase velocity of the wave, becomes slower and
on the average give up part of their energy to the wave and cause
the amplitude of the wave grow. On the other hand, the O VI ions
with velocities slower than the phase velocity of the wave acquire
acceleration, gain energy at the expense of the wave energy.

If the streaming bi-Maxwellian velocity distribution function has a
negative slope then the number of particles causing the wave damping
is greater than the number of particles causing the wave growth. In
this case, the net effect of the ICR process is the transfer of
energy from ICWs to O VI ions.

\section*{Acknowledgments}

Our special thanks go to the referee for his/her valuable
suggestions and B. Kalomeni for reading the manuscript. SD
appreciates the support by the Turkish Academy of Sciences
(T{\"U}BA) Doctoral Fellowship. This study is a part of PhD project
of SD. We dedicate this study to the honorable scientific and
organizational effort given to the Turkish astronomy by Prof. Dr.
Zeki Aslan.

\bibliographystyle{model2-names}

\end{document}